# A study of localisation in dual phase high-strength steels under dynamic loading using digital image correlation and FE analysis

V. Tarigopula[*,1], O. S. Hopperstad[1], M. Langseth[1], A. H. Clausen[1] and F. Hild[2]

[1]Structural Impact Laboratory (SIMLab),
Department of Structural Engineering,
Norwegian University of Science and Technology
NO-7491 Trondheim, Norway

[2]Laboratoire de Mécanique et Technologie (LMT-Cachan),
Ecole Normale Supérieure de Cachan / CNRS-UMR 8535 / Université Paris 6,
61 Avenue du Président Wilson,
F-94235 Cachan Cedex, France

[*]Corresponding author. Tel.: + 47 73 59 46 90; fax: + 47 73 59 47 01.

*E-mail address*: venkatapathi.tarigopula@ntnu.no (V. Tarigopula).

**Abstract**

Tensile tests were conducted on dual-phase high-strength steel in a split-Hopkinson tension bar at a strain-rate in the range of $10^2 - 10^3$ /s and in a servo-hydraulic testing machine at a strain-rate between $10^{-3}$ and $10^0$ /s. A novel specimen design was developed for the Hopkinson bar tests of this sheet material. Digital image correlation was used together with high-speed photography to study strain localisation in the tensile specimens at high rates of strain. By using digital image correlation, it is possible to obtain in-plane displacement and strain fields during non-uniform deformation of the gauge section, and accordingly the strains associated with diffuse and localised necking may be determined. The full-field measurements in high strain-rate tests reveal that strain localisation started even before the maximum load was attained in the specimen. An assessment of the validity of an elasto-viscoplastic constitutive model for the dual-phase steel is provided in terms of its ability to predict the observed stress-strain behaviour and strain localisation. Numerical simulations of dynamic tensile tests were performed using the non-linear explicit FE code LS-DYNA. Simulations were done with shell (plane stress) and brick elements. Good correlation between experiments and numerical predictions was achieved, in terms of engineering stress-strain behaviour, deformed geometry and strain fields. However, mesh density plays a role in the localisation of deformation in numerical simulations, particularly for the shell element analysis.

# 1 Introduction

The high-strength and good formability characteristics of high strength steels when compared to conventional grades make them attractive in applications involving high rates of loading combined with a demand for low weight. Typical examples include light-weight protective systems (Børvik et al., 2006), crashworthiness of automotive and aerospace structures (Tarigopula et al., 2006; Galvez et al., 2006), and high-speed machining. The mechanical behaviour of materials at high strain-rates is considerably different from that observed at quasi-static loading because of the strain rate sensitivity of the material and propagation of stress waves. A standard way of characterizing a material over a wide range of strain-rate is by means of its stress-strain response. In general, the stresses and strains are estimated from the load and displacement measurements by implicitly assuming uniform fields. The data obtained through the stress-strain curve thereby has certain limitations in terms of its validity beyond the point of uniform deformation, where the strains start to localise. Under these conditions, conventional strain measurements will not represent the true behaviour of the material and can only give an indication about an average strain. Furthermore, in high strain-rate tension tests, the onset of necking seems to be a function of the loading rate (Fellows and Harding, 2001), and accurate local measurements of strain localisation are important in studies of dynamic ductile failure.

A vast amount of research has been carried out on localisation phenomena experimentally, theoretically, and numerically (Zener and Holloman, 1944; Noble et al., 1999; Wattrisse et al., 2001; Hopperstad et al., 2003; Stepanov and Babutskii, 2005; Kajberg and Wikman, 2007). In a quasi-static test, it is possible to measure the area reduction in the necked region continuously during deformation with mechanical equipment. However, this method is not suitable for high strain-rate tests. Hopperstad et al. (2003) studied the combined effects of strain rate and stress triaxiality on the deformation behaviour of structural steel using optical measurements. Noble et al. (1999) monitored the nature of localisation at high strain-rates using high-speed photography and then comparisons were performed between the numerically predicted and experimentally observed specimen geometry. However, it is so far not possible to accurately measure the extent of localisation from high speed images of the deformed specimen. It appears that the study of localisation of materials subjected to high strain-rates is still in the early stages due to lack of measurement techniques to obtain accurate field data in the localised region.

In order to obtain more meaningful measurements in the localised region, a robust full-field measurement method is required. Optical measurement methods including laser interferometry, speckle photography and image correlation methods provide promising alternatives. In particular, digital image correlation (DIC) has become popular for full-field measurements in problems related to solid mechanics (Sutton et al., 2000). The advantage of this method is its simplicity and that it sometimes avoids difficult interpretations of interferometric fringes (Hung and Voloshin, 2003). Accordingly, the method has been widely applied to different situations, such as studies of the strain localization phenomena that occur during the tension of thin, flat steel samples, analysis of multi-axial behaviour of rubber-like materials, and microscopic examinations of the fracture processes in concrete (Chu et al., 1985; Choi and Shah, 1997; Wattrisse et al., 2001; Hild et al., 2002;). However, digital image correlation (DIC) in conjunction with high speed photography has not been used very often in high-strain rate tension tests. On the other hand, digital speckle photography has been used by some researchers in investigations of dynamic material behaviour (Grantham et al., 2003; Kajberg and Wikman, 2007). One of the aims of the present paper is to assess the performance of DIC for high strain rate tests.

The most widely used method for high strain-rate testing is the split Hopkinson bar method (Kolsky, 1949) due to its relative simplicity and robustness. Many recent developments have been achieved in its widespread usage from metals to polymers, round to flat specimens, conventional measurements to optical measurements (Verleysen and Degrieck, 2004). However, there are some inherent gripping artefacts associated with the testing of sheet metals (Huh et al., 2002) in the split-Hopkinson bar. In order to avoid these problems, a new specimen design is proposed in this study and its validity has also been checked with respect to attainable dynamic equilibrium in the specimen during deformation.

In the subsequent numerical studies of dynamic processes it is necessary to adopt a constitutive law that accounts for strain-rate dependence. An elasto-viscoplastic constitutive relation was used in the present study. One ultimate test for any constitutive relation is that it should be able to predict different deformation modes depending of the loading conditions, and a reasonable assessment is the model's ability to predict localisation of plastic deformation at high strain-rates.

The present paper aims at studying the localisation in a very ductile dual-phase high strength steel under elevated strain-rates. Low to medium strain-rate tests were performed on a servo-hydraulic testing machine, while high strain-rate tests were conducted using a Split-Hopkinson Tension Bar (SHTB). Real-time measurements of strain and displacements were obtained at high strain-rates using high-speed photography in combination with digital image correlation. The material parameters in an elasto-viscoplastic constitutive model were identified from data obtained in the tension tests. Subsequently, finite element simulations were carried out by using the non-linear explicit FE code LS-DYNA (Hallquist, 1998) and the predicted behaviour of localisation was validated against experimental results.

## 2      Experiments
*2.1      Material*

The material used in the investigation was the cold reduced dual phase high-strength steel alloy, DP800 (C: 0.12 %, Si: 0.20 %, Mn: 1.50 %, P: 0.015 %, S: 0.002 %, Nb: 0.015 %, in terms of % wt) with a minimum yield stress of 500 MPa. This steel is subjected to special heat treatment in the continuous annealing line, which produces a two phase structure, namely ferrite that imparts unique forming properties, and martensite that explicates the strength. All specimens in this study were taken from a sheet with thickness of 1.5 mm and the corresponding variation along the sheet was less than 2 %. In a previous study by Tarigopula et al. (2006), the material was found to be almost isotropic. Thus, it is not essential to carry out tensile tests in different orientations with respect to the rolling direction. It is necessary, however, to perform the tests over a wide range of strain rates in order to identify the parameters of the elasto-viscoplastic constitutive model which is to be applied in the numerical simulations. These tests will also allow observation of the localisation phenomenon under dynamic loading conditions.

All tensile tests were performed at room temperature at a strain-rate in the range of $10^{-3}$ to 600 /s. The enforced strain-rate range is similar to those observed in most practical crash situations, and hence understanding of the properties of the material at these rates is important for safety reasons. In view of the huge variation in applied strain rate levels, different experimental techniques were employed to attain the required strain-rate. The experimental programme comprised 7 tests at different strain rates; 3 tests with a servo-hydraulic testing machine, and 4 tests in a split-Hopkinson tension bar. For the latter, full-field measurements were carried out by using digital image correlation (DIC). Prior to each of the seven tests, width and thickness of the specimens were measured.

*2.2     Tensile tests at high rates of strain*

At high strain rates, from 100 to 600 /s, a Split-Hopkinson tension bar (SHTB) was used (Clausen and Auestad, 2002) to determine the dynamic properties. In SHTB as illustrated in Fig. 1, the specimen was sandwiched between the two long bars of diameter 10 mm (i.e., between C and D), namely the incident (ABC) and output (DE) bars. A stress wave was introduced by clamping the incident bar by means of a friction lock at B, and thereafter pre-stressing part AB of the bar. A strain gauge at position ① was used to monitor the level of the force tension force $N_0$. By releasing the lock at B, a uniaxial elastic tension wave propagates along the incident bar and reaches the specimen, and thereafter the wave is partially reflected and transmitted. These incident, reflected and transmitted pulses were, according to standard procedures for split-Hopkinson bars, measured with strain gauges mounted at positions ② and ③ on the incident and out-put bar. The signals from the strain gauges were monitored and acquired by a personal computer. The transmitted part of the wave accounts for a measure of the load on the specimen, while the reflected part gives a measure of the specimen deformation (Kolsky, 1949; Harding et al., 1960). The strain, stress and strain-rate characteristics in the specimen were calculated from the signals by using standard equations for a Split-Hopkinson bar (Clausen and Auestad, 2002; Tarigopula et al., 2006).

The reliability and validity of the results obtained from Split-Hopkinson tension bar are dependent on the obtainable dynamic equilibrium in the specimen during testing. Dynamic equilibrium ensures an approximately constant strain-rate over most of the test duration or specimen deformation (Song and Chen, 2004). On the other hand, dynamic equilibrium in the specimen is a function of several factors, including transducer properties, pulse shaping, longitudinal wave dispersion and impedance mismatch of the SHTB bars with the specimen (Huh et al., 2002). Specimens are usually designed to be short, so that the transit time for a transmitted pulse is small. This eventually satisfies the dynamic equilibrium condition.

The actual bar has threaded holes with diameter 5 mm at points C and D in Fig. 1. Thus, it is most convenient to test circular specimens with this set-up. This requires, however, that the thickness of the material under investigation is at least 5 mm. The sheet material in this study is 1.5 mm thick. Therefore, a fixture arrangement between the sheet specimen and the M5 threaded holes of the bars had to be designed. These factors led to the advanced design of the specimen and the grips pictured in Fig. 2. The specimen with a uniform gauge length of 5 mm and width of 3 mm, as depicted in Fig. 2a, was used for all tests. The specimen grips have a diameter of 10 mm, which is the same diameter as the bar, and were provided with extensions of 5 mm in diameter in addition to the slot for the specimen, see Fig. 2b. Special care was taken to ensure that the registrations in gauges ② and ③ in combination with one-dimensional wave propagation theory are representative for the response of the test coupon. Thus, any undesirable deformation in the fixture had to be avoided.

First, the two ends of the specimen are glued to the two grips by using Araldite, and this glue binds the specimen to the grips so that it prevents the specimen from sliding. Thereafter threads M5 were machined on the extensions of both the grips which also contained the specimen in the slit. Hence, the specimen was mechanically connected via threads to the M5 holes in the SHTB bars (Fig. 2c), and the glue provides a further reinforcement of this joint. As a result, the fixture will work as an extension of the bar and thereby the impedance mismatch between SHTB bars and the specimen was minimized. Thus, the wave propagation was not disturbed. Finally, the specimens were coated with black and white spray paints to produce a random pattern at the object surface. This random texture facilitates the measurement of the displacement fields through so called correlation techniques.

High speed photography was used to study the deformation of the specimens up to fracture at high rates of strain. A FASTCAM ultima APX 120KC digital CMOS camera, manufactured by Photron, was used to acquire digital images during tests. The advantage of this camera is that it supports full-frame transfer of image without space between pixels. Thus, it gives better quality images so that it eliminates the registration problems between images when applying correlation techniques. The lens of the camera is positioned in front of the specimen at a distance of approximately 200 mm and its axis remains perpendicular to the surface of the specimen. An external flash with the help of a standard illumination system was used to focus the light intensity on the investigated zone of the object surface. In the SHTB tests, a rate of 37500 frames (i.e. an inter-frame time of ~27 µs) was used, yielding a picture with a digitization of 8 bits and a resolution of $512 \times 128$ pixels. It should be noted that in high-speed camera applications, an increase of frame rate will in general reduce the resolution. The selected frequency and resolution is therefore a compromise.

The flash light and the high-speed camera were fired by the time the wave triggered at strain gauge ② in the SHTB bar. Typically, 40 to 70 images were captured for each test from the start of loading to the point at which fracture occurred. The number of pictures was dependent on the strain-rate, as the duration of the SHTB test is inversely proportional with strain-rate.

*2.3    Tensile tests at low rates of strain*

At low to medium strain rates, from $10^{-3}$ to $10^0$ /s, a conventional servo-hydraulic testing machine was used under displacement control, and longitudinal strains were measured with an Instron 2620-601 one sided extensometer over a 3.75 mm gauge length. Same specimen geometry (Fig. 2a) was used for low rates of strain as that of high rates of strain. The displacements were controlled at constant rate and increased from 0.3 to 300 mm/min, prescribing average strain-rates of order $10^{-3}$, $10^{-2}$, and $10^0$ /s, respectively. Additionally, the displacement of the moving crosshead and the applied force were measured by the testing machine. Here, no video measurements were taken. Data were acquired at varying sampling rates in regard to applied strain-rate using an Instron multi-axis software (MAX) running under a Microsoft Windows environment.

## 3    Digital Image Correlation (DIC)

*3.1    Technique*

Digital image correlation is an optical method to measure displacement fields (or strain fields) on an object surface by comparing pictures of the object surface at different states. One state is recorded before loading, i.e. the reference image, and the other states are subsequent images of the deformed object. The displacement field of a planar object has two in-plane components, say $u$ and $v$, and one out-of plane component, $w$. The two in-plane displacement components are directly computed by the digital image correlation. Subsequently, the displacement gradients (or the strain tensors) are derived by space differentiation of the displacement field data.

*3.2  Displacement and strain measurements*

To determine the displacement field on the object surface of the deformed image with respect to a reference image, the region of interest, $ROI_{ref}$, is partitioned into a large number of sub-images and each one is referred to as a zone of interest (ZOI). In this process of evaluating displacements, the correlation principle is used to compare the ZOI in both images (Fig. 3). The method compares the gray value pattern in small sub-images and attempts to evaluate a correlation value using standard procedures of cross-correlation algorithm (Chu et al., 1985; Bruck et al., 1989; Wattrisse et al., 2001), and this value indicates the level of resemblance between two images. One difficulty, however, is the accurate matching of zones of interest which is necessary for a good evaluation of the displacement prior to further refinement. In the present case, a "finite element" correlation algorithm is used (Besnard et al., 2006).

To measure large displacements and strains, coarse-graining technique (Hild et al., 2002) is used to get an evaluation of displacements. As a consequence, the centres of the ZOI in the deformed object surface are moved by the initial estimations and then additional cross-correlation is performed to obtain estimations for subpixel displacements. Here, a new methodology is used to estimate displacement fields from any two selected images captured at different loading stages. The displacement field is decomposed into continuous finite elements (Q4) so that the measured displacement field is consistent with FE analysis computations. It is appropriate to choose either a square or rectangular shape for each Q4 element since the image is divided in to pixels. This laid the foundation for choosing linear shape functions, which are particularly attractive because of the interaction they provide between the measurement of the displacement field by the correlation method and the predicted displacement field by numerical modelling. Last, the obtained subpixel displacements thereafter can be corrected for, if necessary, by using the shift/modulation properties of Fourier transforms. Further details regarding this approach are provided by Besnard et al. (2006).

The evaluated displacement field contains all the information necessary for calculation of the in-plane normal and shear strain components. In order to correspond with the numerical predictions and also because of large displacements, logarithmic strains are used to represent the strain state. These finite strain measures are described in detail in Hild et al. (2002).

*3.3    Out-of-plane displacements*

There are some underlying assumptions (Chu et al., 1985) when using digital image processing: first, it is assumed that the out-of-plane displacement does not influence the in-plane deformation (which is not generally true); second, the gradients of the out-of-plane displacement are assumed to be negligible when compared with the gradients of the in-plane displacements. In the present study, the lens of the camera is placed at a stand-off distance of about 200 mm from the object surface, thus the effect of the out-of-plane displacements on the in-plane displacement of the characteristic intensity surfaces is likely to be negligible.

*3.4    Measurement uncertainties*

It is important to assess the accuracy of the experimental results obtained by using the correlation algorithm. The displacement uncertainty is estimated by considering the recorded picture of one representative test and applying artificially known displacements. In this way, the true performance of the algorithm is predicted with an actual picture. The displacement uncertainty is estimated when the correlation parameters are modified. In particular, the effect of the ZOI size ($l$) on the displacement uncertainty has been investigated. A constant displacement varying between 0 and 1 pixel, with an increment of 0.1 pixel is applied artificially by using the shift/modulation property of Fourier transforms. The average difference gives an indication of the error and the standard deviation gives an evaluation of the corresponding uncertainty. The quality of the estimate is characterized by the standard uncertainty, $\sigma_u$, which is defined as the mean of the standard displacement uncertainties. Fig. 4 presents the standard uncertainty as a function of the ZOI size, which shows that the larger the ZOI size, the smaller the uncertainty. This means that the displacement uncertainty and the corresponding spatial resolution are the result of a compromise. The preferred size of the sub-images $12 \times 12$ pixels is found sufficient to provide accurate results for the chosen strain-rate specimen. It provides a displacement uncertainty of the order of 1 centi-pixel, and a strain uncertainty of the order of $10^{-3}$.

**4      Experimental results**

Seven tests were performed; three at low rate and four at high-rate, and the corresponding material data are listed in Table 1. The tests in the servo-hydraulic machine are referred as "lsr", while the tests in the SHTB are referred as "hsr". Both engineering and true mean strain-rates for each test are given in Table 1. They were estimated as the average value of the strain rates recorded from each test.

As discussed earlier, results from the SHTB technique are valid only in the region where the specimen is in dynamic equilibrium. Fig. 5a shows the measured signals obtained from the strain gauges of the SHTB for a representative test, hsr03, at a strain-rate of 445 /s. This figure indicates that equilibrium has been attained in the specimen without any dispersion effect, but after some plastic deformation has taken place. It should be noted that the data in the linear elastic region of the stress-strain curve comes from the early time when stress is not in equilibrium. Hence, Young's modulus values from the SHTB tests were not taken into account.

The true stress-plastic strain curves until diffuse necking are shown in Fig. 5b. It is observed that the strain-rate has an influence on the strength and the increase of the dynamic flow stress is approximately 12 % within the investigated strain-rate range. Fig. 6 presents the selected sequence of high-speed images that covers the entire test from unloaded specimen (frame 8) to completely broken specimen (frame 45) for the chosen representative test at a strain-rate of ~ 445 /s. From these images, localisation by necking starts between frames 36 and 38, i.e. between 0.75 ms and 0.8 ms from the start of the loading, and thus up to frame 36 the plastic deformation is visibly uniform. Even though diffuse necking starts somewhere in the vicinity of frame 32 (from load-time characteristics) it is not clearly visible in the deformed shape. Subsequently, local necking occurs somewhere between frames 41 and 42, i.e. between 0.88 ms and 0.91 ms. Eventually, the onset of failure is observed close to the centre of the gauge section at around frame 44, i.e. at ~ 0.96 ms. The same trend is observed for all high strain-rate tests with respect to localisation phenomena, i.e. no clearly visible diffuse neck occurs prior to localised necking.

In order to acquire quantitative measurements of specimen displacements and strains from the high-speed photography, digital image correlation was employed. The uncertainty in the displacement measurements for this test is of the order of $10^{-2}$ pixel (see Fig. 4) based on the aforementioned correlation algorithm for the selected ZOI size of 12 pixels. The corresponding mesh for the selected ZOI size is shown in Fig. 7a. First, the displacement field was measured by tracking the position of the contrast change (pixel number) while updating the image. A total of 15 image updates are used for the evaluation and the images are chosen randomly. In the beginning, images are selected at very large intervals and this interval was steadily narrowed down towards the end of the test, due to the fact that the strain values increase more rapidly towards the end of the test compared to the beginning.

Subsequently, the strains were computed at the centre of the specimen gauge section (Fig. 7b) based on known $u$ and $v$ displacement components. The zone was selected (Fig. 7b) over the area in which localisation is confined. Thus, the computed local strains in the selected zone (shown in Fig. 7b) give an indication about the extent of localisation, but may not exactly represent the failure strain. This is due to the fact that the selected zone is concentrated over the wider area and not focussed on a point where the onset of failure occurs. Plots of the longitudinal logarithmic strain ($\varepsilon_x$) across the centre of the specimen as a function of time for two representative strain-rate tests hsr02 and hsr03 corresponding to 360 /s and 445 /s, respectively, are shown in Fig. 8. In order to obtain an estimate of the failure strain, the selected zone is refocused on the area at which a localised neck was formed. An average logarithmic failure strain of approximately $50 \pm 5$ % was obtained for all high rates of strain. In addition, microscope measurements were performed on the ruptured post-test specimens in order to check the fracture strain, and these strains found from microscope measurements agree quite well with the predicted fracture strains by DIC.

## 5    Constitutive model and parameter identification

### 5.1    *Constitutive model*

In order to perform finite element simulations of the dynamic tensile tests, an appropriate constitutive model is required that represents the strain-rate sensitivity of the material observed in the experiments. Hence, it is necessary to consider elasto-viscoplastic constitutive relations such as those proposed by Johnson and Cook (1983) or Zerilli and Armstrong (1987), which are often used to describe the mechanical response of metallic materials at elevated rates of strain. In this work, an elasto-viscoplastic constitutive model, which is described in detail by Reyes et al. (2006), was adopted in the numerical predictions. This model includes isotropic elasticity, anisotropic yield criterion, associated flow rule, combined isotropic and kinematic hardening, and strain-rate effects. This model is a standard model (*MAT_135) in LS-DYNA. In the present study, plastic anisotropy and kinematic hardening were neglected, since the material is more or less isotropic (Tarigopula et al., 2006) and strain path change or reversed loading is not considered.

*5.1.1 Constitutive relation*

The effective stress ($\bar{\sigma}$) is expressed, assuming non-linear isotropic hardening and isothermal conditions, by a modified Voce rule in the form

$$\bar{\sigma} = (\sigma_0 + \sum_{i=1}^{2} Q_i(1 - \exp(-C_i \bar{\varepsilon})))(1 + \frac{\dot{\bar{\varepsilon}}}{\dot{\varepsilon}_0})^q, \qquad (1)$$

where $\sigma_0$, $Q_i$, $C_i$ and $q$ are material constants, $\bar{\varepsilon}$ is the effective plastic strain, $\dot{\bar{\varepsilon}}$ is the effective plastic strain-rate and $\dot{\varepsilon}_0$ is a user-defined reference strain-rate. The first factor in the equation governs strain hardening, while second factor defines the rate dependent behaviour (viscoplasticity). The involved material constants are determined from the seven tensile tests.

*5.1.2 Yield criterion*

An isotropic yield criterion with high exponent is adopted in the simulations (Hershey, 1954), and the effective stress is then defined as

$$\bar{\sigma} = \left[\frac{1}{2}\left\{\left|\sigma_1 - \sigma_2\right|^m + \left|\sigma_2 - \sigma_3\right|^m + \left|\sigma_3 - \sigma_1\right|^m\right\}\right]^{\frac{1}{m}}, \qquad (2)$$

where $(\sigma_1, \sigma_2, \sigma_3)$ are the principal stresses. In the case of plane stress, the principal stress in the thickness direction of the sheet equals zero: $\sigma_3 = 0$.

The recommended values for exponent $m$ are 6 for bcc materials and 8 for fcc materials (Logan and Hosford, 1980). It is noted that for $m = 2$ the criterion coincides with the von Mises yield criterion, while in this study $m = 6$ was used.

*5.2    Identification of material parameters*

Identification of material parameters requires representative calibration tests. The selection of appropriate tests, to a large extent, would depend on the complexity of the test as well as its range of application. The material parameters in the chosen constitutive model were identified from uniaxial tensile tests over a wide range of strain-rates.

The model, which accounts for elasto-viscoplasticity with non-linear isotropic hardening, possesses seven parameters ($\sigma_0$, $Q_1$, $C_1$, $Q_2$, $C_2$, $q$, $\dot{\varepsilon}_0$) as indicated in Eq. (1). The strain hardening Voce parameters $\sigma_0$, $Q_i$ and $C_i$ ($i = 1, 2$) in Eq. (1) were derived by curve fitting the equation to the quasi-static true stress-plastic strain curve at a strain-rate of $8 \times 10^{-4}$ /s, which is chosen as the reference curve. Fig. 9a shows that this selected model describes the experimental behaviour very well before necking, and the corresponding calibrated parameters are presented in Table 2.

In order to determine the strain-rate dependence, it is necessary to know the relation between the strain rate, $\dot{\bar{\varepsilon}}$, and the effective stress, $\bar{\sigma}$, at a particular level of the plastic strain, $\bar{\varepsilon}$. The flow stresses with respect to the strain-rates are displayed in Fig. 9b, plotted on a semi-log scale at 8 % and 10 % plastic strain levels. The viscoplastic parameters were found by using least squares fit between the model represented by Eq. (1) and the experimental data. The identified model is in good agreement with the experimental behaviour when $q = 0.0175$ and $\dot{\varepsilon}_0 = 0.78$ /s at a plastic strain of 10%, and the same model parameters also correspond well with the experimental behaviour at a plastic strain of 8 %, see Fig. 9b.

## 6      Numerical simulations

The explicit non-linear finite element code LS-DYNA 970 (Hallquist, 1998) was used for performing the simulations of the experimental tests. A finite element model of the dynamic tensile specimen was created by using the mesh generator program TrueGrid (2001). The simulations were performed using the material model (*MAT_135) as presented in Section 5 and the corresponding model parameters are gathered in Table 2.

Four-node quadrilateral Belytschko-Tsay shell elements were used for the plane stress modelling of the entire dynamic tensile specimen. Furthermore, the elastic steel fixtures were also modelled along with the bars of the SHTB partially by using 3D brick elements as depicted in Fig. 10a. This was done primarily because the fixtures were connected along the extensions of the specimen and bonded together by glue and threads in the real test. A tied surface-to-surface contact was introduced to model the interaction between the specimen and fixtures. The contact between the bars and the fixtures was also modelled using surface-to-surface contact. A velocity boundary condition through a smooth cosine curve (Fig. 10b) was applied at the left rigid bar while the right bar was fixed. The employed velocity in the simulations was obtained from the experimental measurements, and this is the average relative velocity (i.e. $\Delta v = \dot{\varepsilon} L_0$, where $\dot{\varepsilon}$ is the true strain-rate and $L_0$ is the specimen gauge length) in the specimen during the test. For example, the velocity applied for the chosen representative test at 445 /s was 2.225 m/s. The specimen dimensions were identical to those of the experimental specimens with a gauge length of 5 mm and the sheet thickness of 1.5 mm. An element size of 0.25 mm × 0.125 mm was found reasonable in the gauge area to predict localisation, creating a total of 4960 shell elements. Stiffness-type hourglass control was used to eliminate zero energy modes.

Fig. 11 outlines the deformed shape of the FE mesh at five stages during the analysis of Split-Hopkinson bar test at 445 /s. It is seen that a neck was formed at a point close to the centre of the gauge section, which is similar to the experimental observations, as shown in Fig. 6. A comparison between the experimental and predicted strain field $\varepsilon_x$ is presented in Fig. 11, and it is seen that the fields are in good agreement with each other.

## 7     Comparison and discussion

Experimental stress-strain behaviour and measurements of strain localisation in dynamic tensile tests are compared with numerical predictions based on the constitutive model defined by Eq. (1). Isothermal conditions are assumed in the FE simulations. In general, plastic work is converted into heat and the temperature of the material rises when a material is tested at high rates of strain (Noble and Harding, 1994). Adiabatic conditions prevail, and the rate of temperature increase is usually defined as (Bammann et al., 1993; Børvik et al., 2001)

$$\dot{T} \approx \chi \frac{\boldsymbol{\sigma}:\mathbf{D}^p}{\rho C_p}, \qquad (3)$$

where $\boldsymbol{\sigma}$ is the stress tensor, $\mathbf{D}^p$ is the plastic rate-of-deformation, $C_p$ is the specific heat at constant pressure (~ 452 J/kgK for steel) and $\rho$ is the mass density (~ 7850 kg/m$^3$), and the Taylor-Quinney empirical constant $\chi$ is often assumed with the value 0.9. It was found based on Eq. (3) for the chosen test at 445 /s that the rise in temperature is approximately 60 K in the region of large strains. This moderate temperature increase is considered to be negligible and hence the assumed isothermal conditions in the numerical model should provide good predictions.

From Fig. 12a & b it is observed that the shape of the measured and predicted deformed geometries at necking are in good agreement, and in both the cases the neck was formed at a point close to the centre of the gauge area. Fig. 12c illustrates the correlation between the experimental and predicted engineering stress-strain curve for a split-Hopkinson bar test at 445 /s. It is apparent from the figure that the numerical predictions are in good agreement with experimental measurements, except the slight discrepancy for small strains. This discrepancy is attributed to the initial non-uniform stress state in the specimen during testing. Moreover, it is not virtually possible to obtain information about strain localisation from measurements of force and deformation. The digital image correlation method together with high-speed photography provided the useful observations of the start of necking, and the displacement and strain fields in the localised zone at high strain-rates. Digital image correlation measurements for the strain-rate test, hsr03, at 445 /s (Fig. 12d) show that the material reaches a maximum load at a local engineering strain of approximately 0.18 (i.e., ~ 0.64 ms or around frame 32 in Fig. 6). In the SHTB measurements, it is observed from Fig. 12c that the engineering strain at the maximum load is significantly lower and approximately equal to 0.13. This is due to the fact that the strain has already started to localise at the point of maximum load, and while the local behaviour is well captured with the DIC measurements, the SHTB measurements merely give the overall response. Furthermore, the characteristic feature of localised necking is clearly apparent roughly at 0.88 ms (i.e., around frame 38 in Fig. 6) in the tests. The intermittent frames between 32 and 38 demonstrate a localisation mechanism exhibiting a gradual transition from uniform deformation to necking. This means that a stable deformation, in the sense that the load decreases slowly, is still possible beyond the maximum load in the high strain-rate tests.

Moreover, numerical simulations validated the above experimental observations. As seen from Fig. 12d, a very good agreement prior to localisation was observed between the digital image correlation results, in terms of axial logarithmic strains with respect to time, and numerical simulations, while a very small drift is observed thereafter. This drift might be due to inaccuracies in the FE simulation (e.g. the plane stress assumption) or in the optical measurements, or a combination of both. Inaccuracies in the optical measurements might be because of possible aberrations involved in the imaging at different scales, or the large measurement uncertainties associated with large strains. The aberrations could easily be controlled by calibrating the camera before every test (Kajberg et al., 2004). Furthermore, the measurement uncertainties depend on several factors such as correlation accuracy, sub-pixel algorithm, and these issues were addressed in Section 3.1.3. And finally, the finite number of pixels (512 × 128) also implies an approximation and as such a potential reason for small variances between FE and optical predictions of the specimen behaviour. One should note that the logarithmic strains were computed (Fig. 12d) from the centre part of the gauge section both from experiments (digital correlation method) and simulations in order to be consistent. Since there is no failure criterion used in the FE simulations, comparisons were made with experiments only up to incipient failure.

In addition, Fig. 13 shows the development of the predicted localisation between the point of maximum load to the onset of local necking. The contours of predicted plastic strains in Fig. 13 distinctly suggest that strain localisation in the form of diffuse necking has already started at approximately 0.63 ms, and thus localisation started even before the peak load in the specimen at high strain rate tests. This is in accordance with DIC measurements. From Fig. 13, it is also pointed out that after the start of diffuse necking (~ 0.63 ms) the localised plastic deformation is spread uniformly until incipient local necking (~ 0.78 ms – 0.80 ms). Thus, numerical simulations substantiate the experimental observation of stable deformation beyond peak load in the sense that gradual transition from uniform strains to non-uniform strains occurs at high rates of strain. These results are in broad agreement with the findings of Noble et al. (1999) in spite of the different material and specimen geometry.

In order to check the effect of the plane stress assumption, simulations were also carried out using eight-node brick elements. The 3D model of the specimen generated 32640 solid elements with an element size of 0.125 mm $\times$ 0.125 mm $\times$ 0.19 mm in the gauge section. This gives 24 elements across the width and 8 elements over the thickness in the gauge section. Comparisons were made between experiments and numerical simulations using shell and solid elements. Again excellent agreement was observed between experiments and numerical predictions of changes in specimen geometry at the stage of localized necking as shown in Fig. 14a & b. However, Fig. 14c & d indicate that the brick element simulation gives less accurate results than the one with shell elements with respect to experimental results. This conclusion is not generally valid, since the results obtained with the shell model depend strongly on mesh density. Fig. 15 shows that refinement of the mesh (0.125 mm $\times$ 0.125 mm in the gauge section) with shell elements leads to premature localisation, while the brick element model is less sensitive to mesh refinement (0.0625 mm $\times$ 0.0625 mm $\times$ 0.12 mm that gives a 48 elements along the width direction and 12 elements over the thickness). Hence, it is important to choose an appropriate mesh density in the shell model in order to correctly predict strain localisation. Improved convergence properties of shell models with respect to strain localization were obtained by use of non-local plastic thinning (*MAT_NONLOCAL option in LS-DYNA), as shown in Fig. 15a. A non-local approach to plastic thinning stabilizes the overly localised deformation within a prescribed radius of influence surrounding the integration point in the thickness direction of the shell. In this study, the radius of the non-local domain is taken as half of the actual thickness of the sheet (i.e. 0.70 mm). More details about this approach are provided by Wang et al. (2006).

## 8    Conclusions

Quasi-static and dynamic tensile were performed for a high-strength dual-phase steel. High speed photography was used for the dynamic tests to capture the inception and development of strain localisation. A full-field measuring system based on digital image correlation was successfully applied to track the strain field from initial plasticity to fracture, and thus provided some substantial information concerning the localisation and hence the inception of ductile failure. The quality of the frames is sufficient to determine a displacement field by comparing two digital images. The advantage of this novel approach is that it is globally continuous despite the heterogeneous strain fields. The experimental results from split-Hopkinson bar tests indicated an early localisation and gradual transition from uniform strains to non-uniform strains.

Subsequently, the experimental behaviour was modelled using LS-DYNA and an elasto-viscoplastic constitutive model. Simulation of the experiments using estimated parameters showed excellent agreement with experimental results in terms of load-deformation characteristics and mode of deformation. However, selection of appropriate mesh size in the FE simulations with shell elements is very important for accurate prediction of the evolution of localisation. It is transpired from the present study that the suggested digital image correlation method is viable to most practical problems and bridges the gap between experiments and numerical simulations. However, a further development of this method is necessary to extend its application from 2D images to stereovision so that this technique is broadly applicable for crash testing and other dynamic phenomenon. Care should also be taken in improving, if possible, the spatial resolution in optical measurements.

**Acknowledgement**

The present work has been carried out with financial support from the Research Council of Norway through the strategic university programme "Design of Crashworthy Light Structures" and support from the Norwegian University of Science and Technology. Thankful acknowledgement is also made to SSAB for supplying the test material.

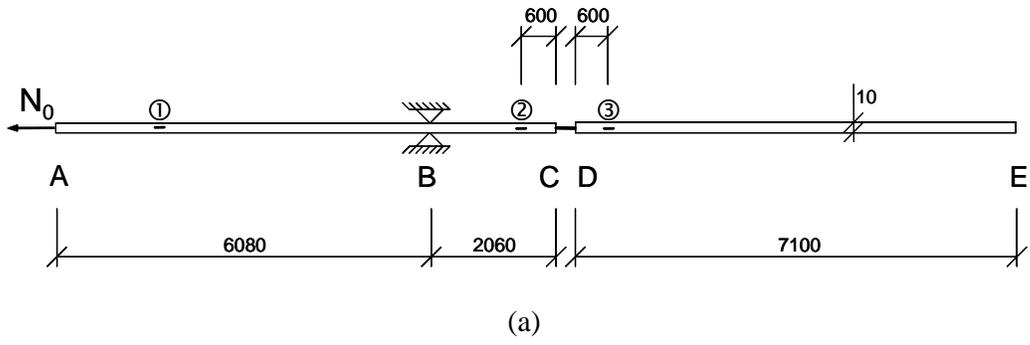

(a)

Fig. 1. Schematic diagram of split-Hopkinson tension bar (SHTB) set-up.

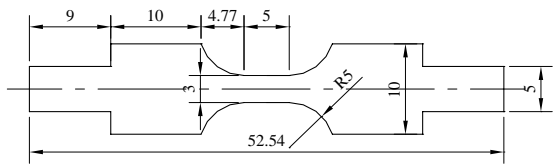
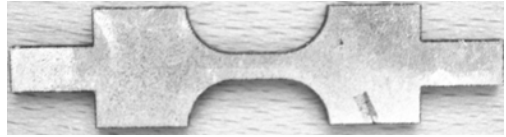

(a)

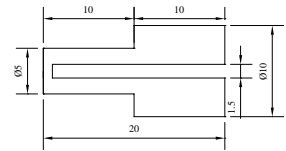
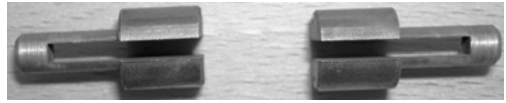

(b)

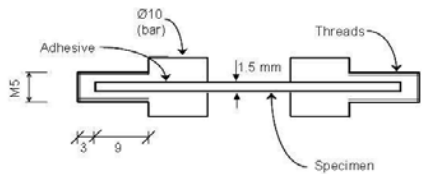
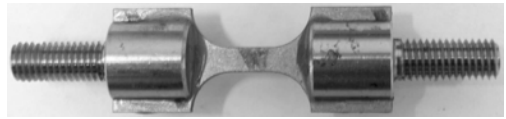

(c)

Fig. 2 (a) Geometry of the specimen, (b) Geometry of the fixture and (c) Assembly of the specimen and fixture

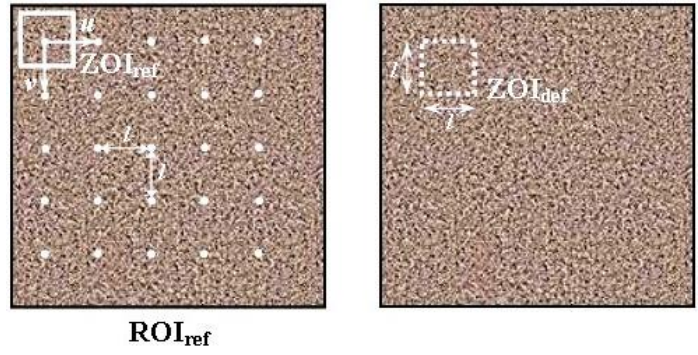

(a)                  (b)

Fig. 3. Correlation of the reference (a) and deformed (b) images ($l$ is the size of sub-image as well as inter-distance between two pixels in the Q4-procedure)

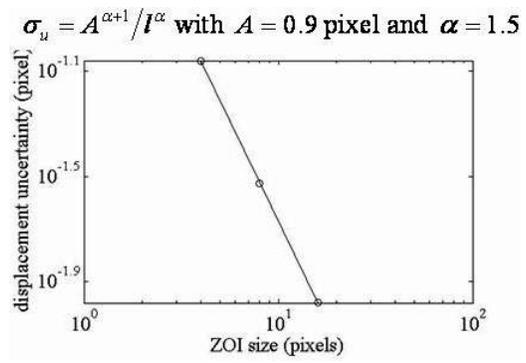

Fig. 4. Displacement uncertainty as a function of the ZOI size for the strain-rate test at 445 /s

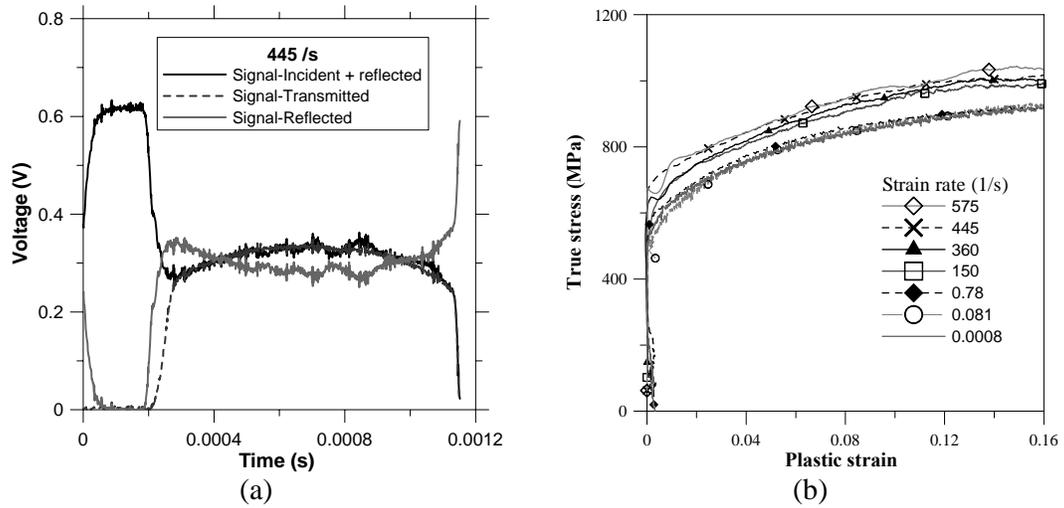

Fig. 5 (a) Voltage signals for a representative test, hsr03, at 445 /s and (b) True stress-plastic strain curves from dynamic tensile tests

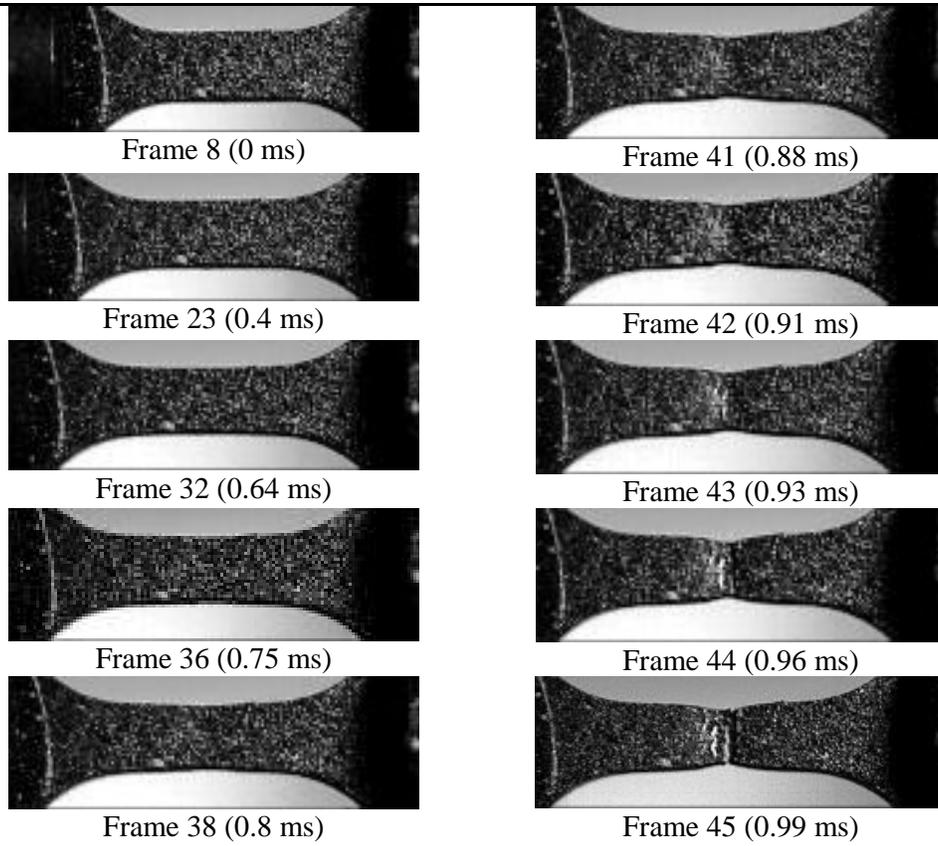

Fig. 6. Sequence of the high-speed images for a strain-rate test, hsr03, at ~ 445 /s (resolution: 512 × 128 pixels, digitisation: 8 bits)

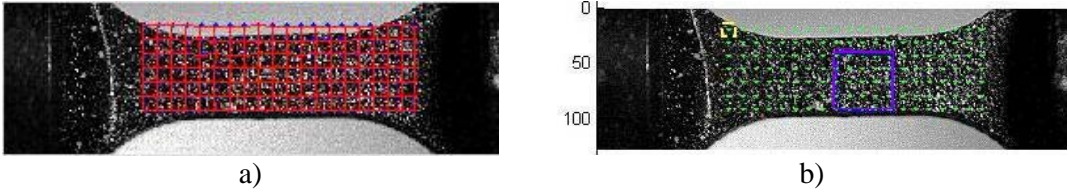

Fig. 7 (a) Mesh for selected ZOI size (= 12 pixels) and (b) Selected zone on virtual grid for the estimation of strains

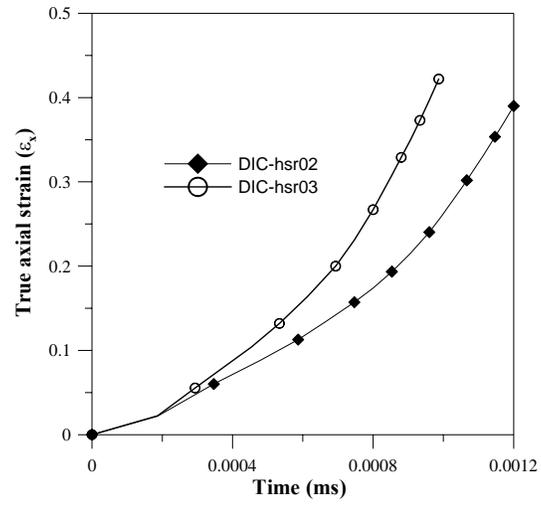

Fig. 8. Digital image correlation measurements: axial logarithmic strain vs. time plot for two strain-rate tests, hsr02 and hsr03.

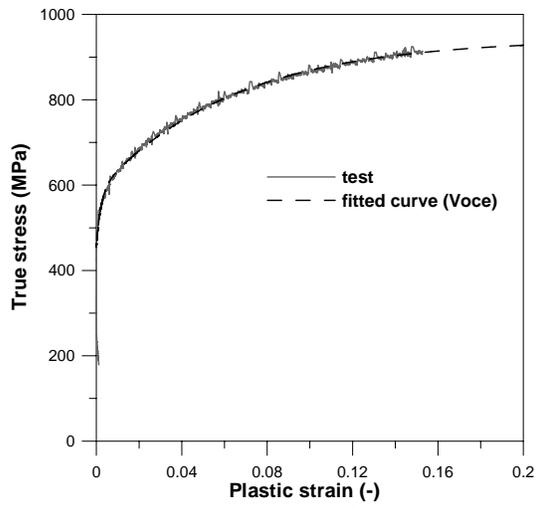 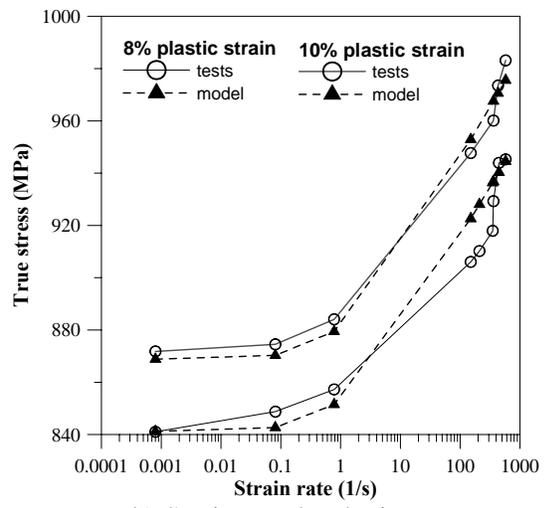

a) Strain-hardening   b) Strain-rate hardening

Fig. 9. Comparison of curves from tests and fitted model, Eq. (1)

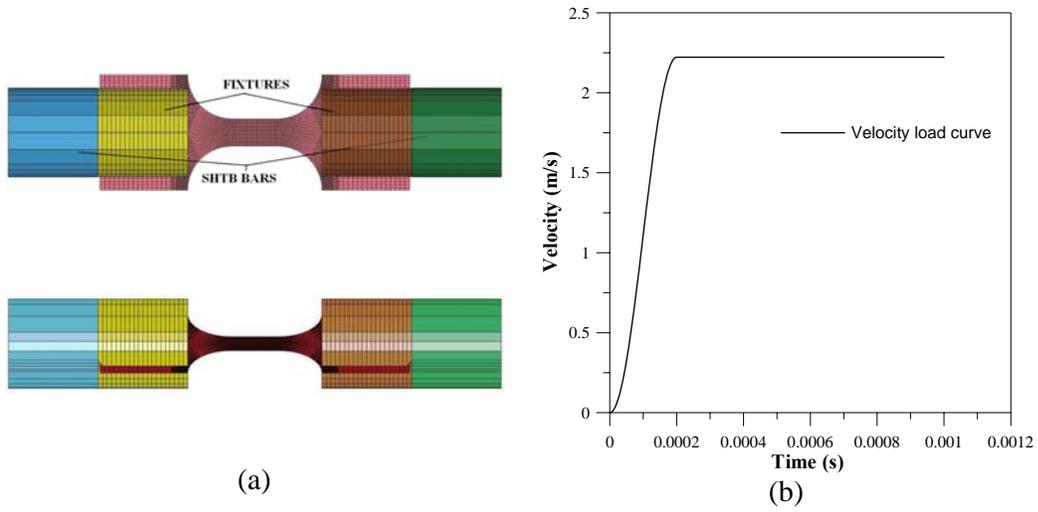

Fig. 10 (a) FE model of the dynamic tensile specimen and (b) load curve used in dynamic tensile simulations for a strain-rate of 445 /s

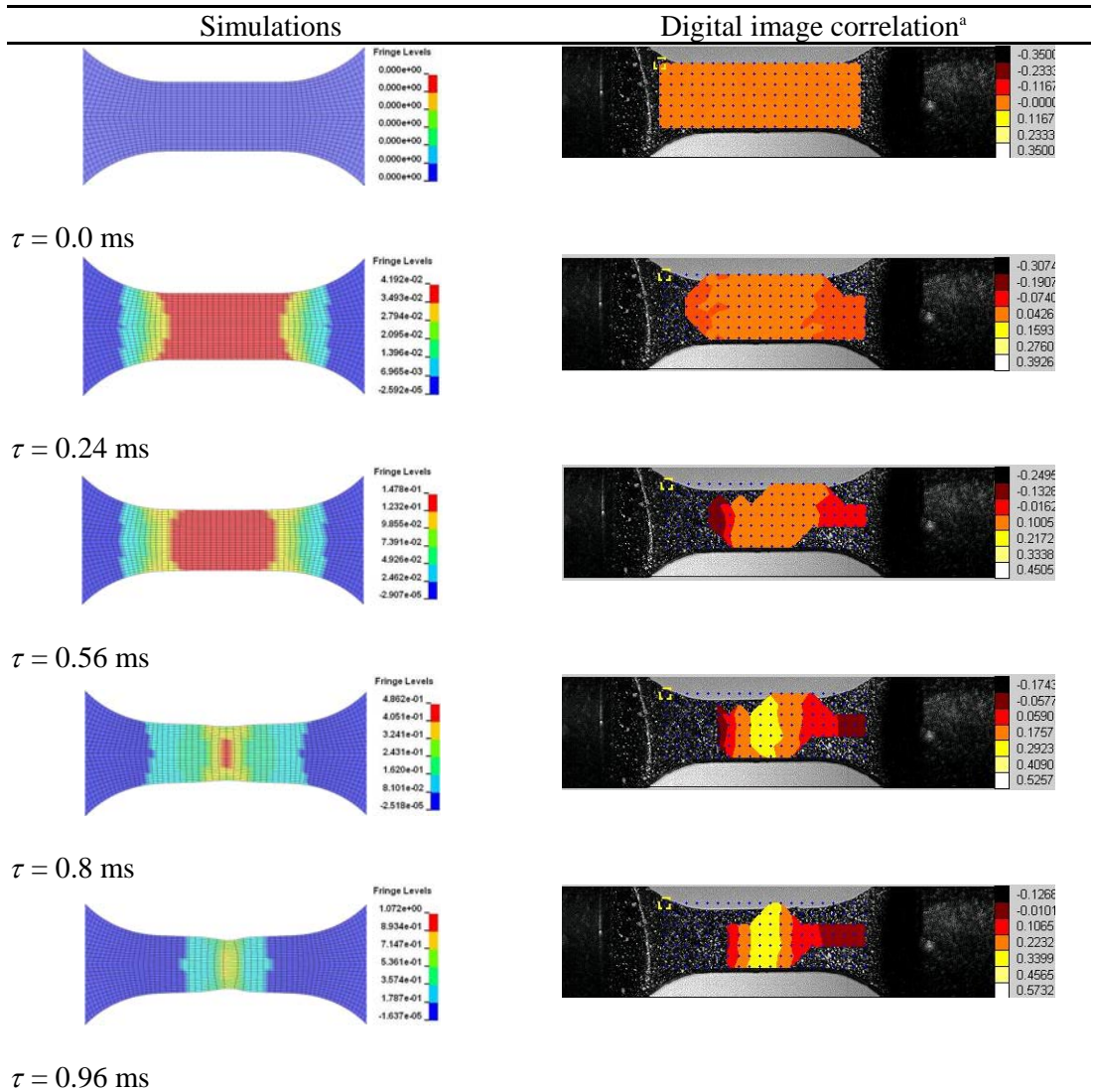

Fig. 11. Selection of axial logarithmic strain component, $\varepsilon_x$, frames at various time rates from simulations and correlation algorithm

[a] The image in the background is always the reference image, while the colour plots represent the deformed strain field

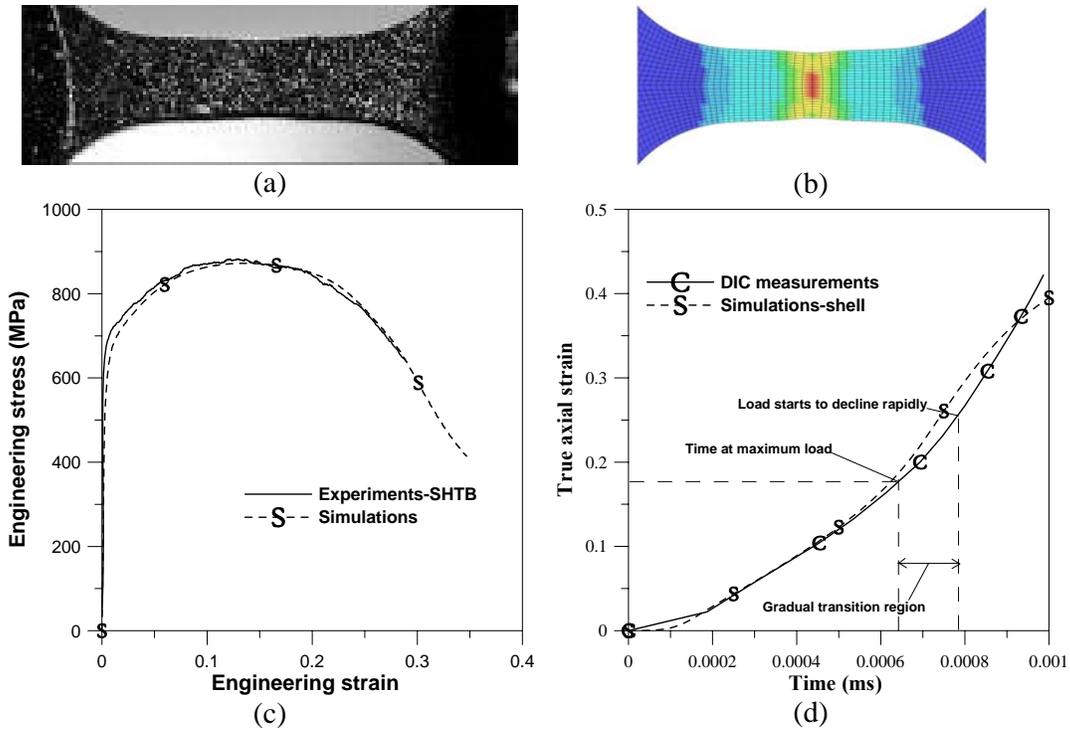

Fig. 12. For a representative test, hsr03, at 445 /s, (a) Test specimen at the onset of necking, (b) Predicted specimen geometry at the onset of necking, c) Comparison of nominal stress-strain curves between experiments and simulations with shell modelling, and (d) Comparison of axial logarithmic strain vs. time plots from experiments and simulations with shell element model

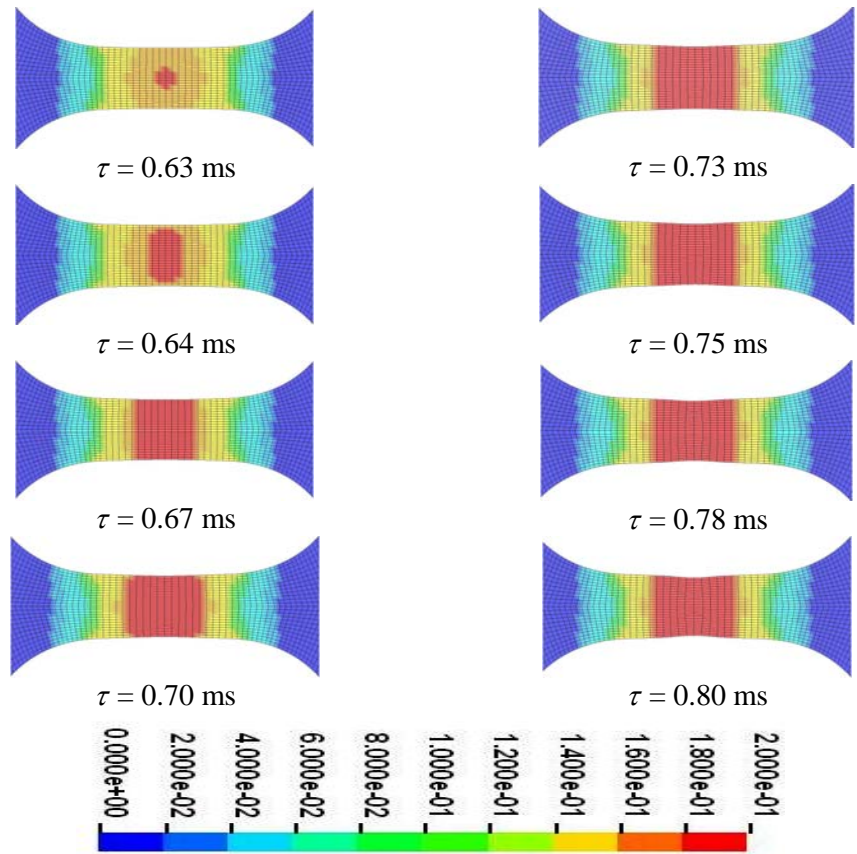

Fig. 13. Sequence of the development of localisation, in the form of plastic strain contours[b], in numerical simulations for a strain-rate at 445 /s

[b] The colour maps of plastic strain were obtained with the fixed fringe range, so that one can clearly see the initiation and propagation of localisation

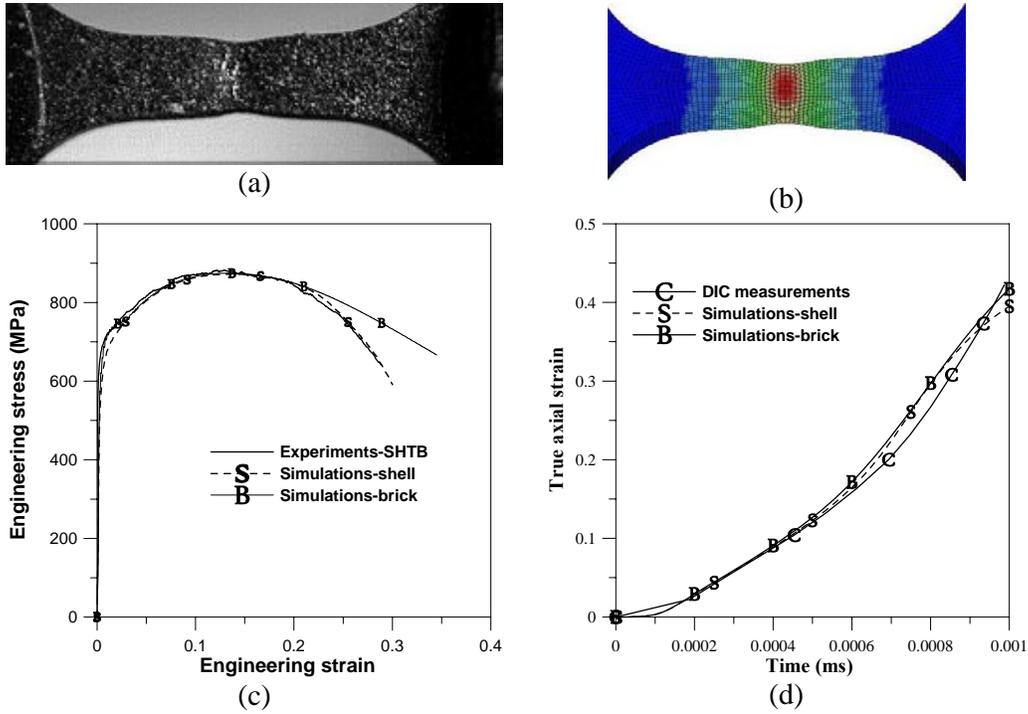

Fig. 14. For a representative test, hsr03, at 445 /s, (a) Test specimen at the stage of localized necking, (b) Predicted geometry using solid elements at the stage of localized necking, (c) Comparison of nominal stress-strain curves between experiments, simulations with shell modelling and simulations with brick modelling, and (d) Comparison of axial logarithmic strain vs. time plots between experiments, simulations with shell modelling and simulations with brick modelling

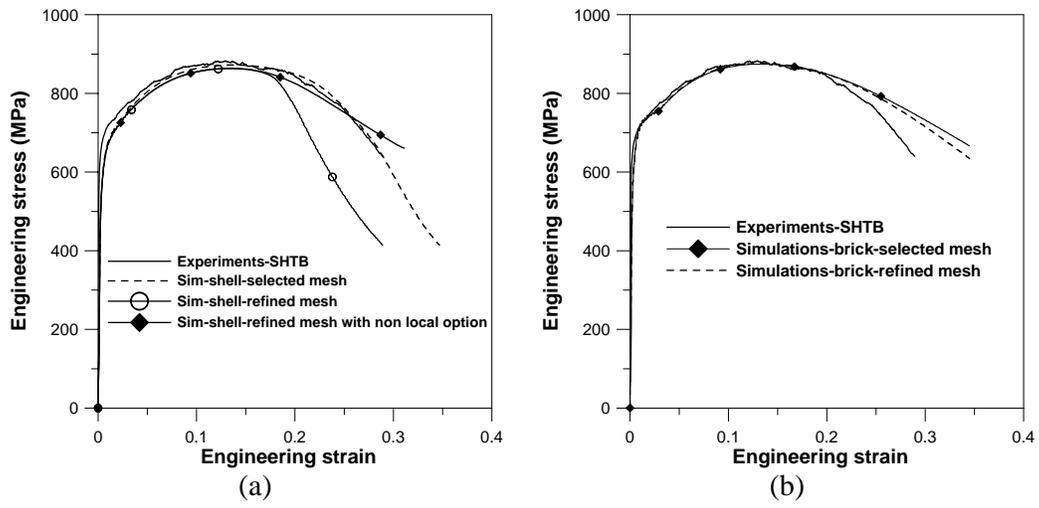

Fig. 15. Influence of mesh density on engineering stress-strain response in (a) shell element model and (b) brick element model

Table 1. Material data for strain-rate tests

| Test | Width (mm) | Thickness (mm) | Strain-rates (1/s) | | Yield stress $s_{0.2}$ (MPa) | Max stress $s_u$ (MPa) | Strain at $s_u$ $e_u$ |
|---|---|---|---|---|---|---|---|
| | | | Engineering $\dot{e}$ | True $\dot{\varepsilon}$ | | | |
| lsr01 | 3.054 | 1.494 | 0.00078 | 0.00070 | 530 | 797 | 0.139 |
| lsr02 | 3.060 | 1.497 | 0.081 | 0.069 | 521 | 801 | 0.144 |
| lsr03 | 3.062 | 1.495 | 0.775 | 0.645 | 550 | 801 | 0.152 |
| hsr01 | 3.075 | 1.502 | 153 | 127 | 584 | 864 | 0.114 |
| hsr02 | 3.064 | 1.494 | 362 | 306 | 644 | 876 | 0.139 |
| hsr03 | 3.069 | 1.506 | 445 | 377 | 689 | 883 | 0.122 |
| hsr04 | 3.065 | 1.495 | 575 | 488 | 659 | 900 | 0.136 |

Table 2. Identified parameters from the experiments for DP800

| $E$ [GPa] | $\nu$ [-] | $\sigma_{0.2}$ [MPa] | $\sigma_0$ [MPa] | $Q_1$ [MPa] | $C_1$ [-] | $Q_2$ [MPa] | $C_2$ [-] |
|---|---|---|---|---|---|---|---|
| 205 | 0.30 | 530 | 450 | 130 | 475 | 360 | 15.8 |

| $\dot{\varepsilon}_0$ [1/s] | $q$ [-] | $m$ [-] |
|---|---|---|
| 0.78 | 0.018 | 6 |